\documentclass[preprint,12pt]{revtex4-2}
\usepackage{array}
\usepackage{graphicx}
\usepackage{dcolumn}
\usepackage{bm}
\usepackage{amsmath}
\usepackage{amssymb}
\usepackage{verbatim}
\usepackage{tabularray}
\usepackage{comment}
\usepackage{siunitx}
\usepackage{mathtools}
\usepackage[english]{babel}

\begin{document}

\title{Magnetoplasma excitations in interacting GaAs disks}

\author{S.~A.~Andreeva}\email{lopatina@issp.ac.ru}
\affiliation{National Research University Higher School of Economics, 101000 Moscow, Russia}

\affiliation{Osipyan Institute of Solid State Physics RAS, 142432 Chernogolovka, Moscow district, Russia}

\author{A.~A.~Gavrilov}
\affiliation{Osipyan Institute of Solid State Physics RAS, 142432 Chernogolovka, Moscow district, Russia}

\author{K.~R.~Dzhikirba}
\affiliation{Osipyan Institute of Solid State Physics RAS, 142432 Chernogolovka, Moscow district, Russia}

\author{A.~S.~Astrakhantseva}
\affiliation{Osipyan Institute of Solid State Physics RAS, 142432 Chernogolovka, Moscow district, Russia}

\author{A.~V.~Shchepetilnikov}
\affiliation{Osipyan Institute of Solid State Physics RAS, 142432 Chernogolovka, Moscow district, Russia}

\author{O.~V.~Orlov}
\affiliation{Osipyan Institute of Solid State Physics RAS, 142432 Chernogolovka, Moscow district, Russia}

\author{V.~V.~Solovyev}
\affiliation{Osipyan Institute of Solid State Physics RAS, 142432 Chernogolovka, Moscow district, Russia}

\author{I.~V.~Kukushkin}
\affiliation{Osipyan Institute of Solid State Physics RAS, 142432 Chernogolovka, Moscow district, Russia}

\date{\today}

\begin{abstract}
We investigate the effect of inter-disk coupling on the magnetoplasmon dispersion in a square lattice of two-dimensional electron system (2DES) disks etched from a GaAs quantum well. Using magneto-optical terahertz (THz) spectroscopy, we track the evolution of the collective modes as disk lattice period is systematically reduced, thereby increasing the coupling strength. At large distances, the system exhibits magnetoplasma modes corresponding to individual excitations in disks. As the inter-disk distance decreases, we observe a modification to magnetoplasma dispersion.

\end{abstract}

\maketitle

\section{Introduction}

Plasmons in two-dimensional electron system (2DES) have long served as a fertile ground for exploring fundamental many-body physics and for developing plasmonic devices in the infrared and terahertz regimes~\cite{ju2011graphene,jin2017infrared}. In the presence of a perpendicular magnetic field, the 2DES supports magnetoplasmon hybrid excitations that couple collective charge density oscillations with cyclotron motion. Their dispersion, which exhibits characteristic branches such as the upper and lower hybrid modes, is highly tunable via the magnetic field strength and the electron density~\cite{PhysRevB.104.195436,kukushkin2003observation}. 

Over the past decade, advances in nanofabrication have enabled the patterning of 2DESs into mesoscopic structures, such as quantum dots, antidots, and disks, where geometrical confinement quantizes the magnetoplasmon spectrum~\cite{bitton2019quantum,hatef2012plasmonic,liu2015electrically}. In particular, arrays of isolated 2DES disks have been extensively studied, revealing size-dependent resonances that can be understood within the framework of edge and bulk magnetoplasmons~\cite{volkov1985theory,volkov1988edge,PhysRevB.28.4875,fei2015edge}.

However, most prior work has focused on the non-interacting or weakly interacting limit, where the distance between disks is large enough to render inter-disk coupling negligible. In this case, the plasma frequency is given by the known relationship:
\begin{equation}
\omega_p (q)=\sqrt{\dfrac{n_s e^2 q}{2m^{\ast} \varepsilon_0  \varepsilon}},
\label{2D}
\end{equation}
where $n_{s}$, $m^{\ast}$ -- 2D electron density and effective mass, respectively, $\varepsilon$ -- effective dielectric permittivity of the environment media. The fundamental plasma mode corresponds to the wave vector $q_{0} = 1.2/R$, where $R$ is a single disk radius~\cite{kukushkin2003observation}. In the presence of a perpendicular magnetic field, as mentioned above, fundamental plasma mode splits into two branches:
\begin{equation}
\omega_{\pm} =\sqrt{\omega_p^{2}+\left(\frac{\omega_c}{2} \right)^{2}} \pm \frac{\omega_c}{2},
\label{disp}
\end{equation}
where $\omega_c = eB/m^{\ast}$ — the cyclotron resonance frequency.

When disks are brought into close proximity, the evanescent electromagnetic fields of their collective modes overlap, leading to hybridization, anti-crossings, and the formation of extended Bloch-like plasmonic bands. This regime is of significant current interest because it bridges the gap between localized plasmon resonances in metamolecules and propagating surface magnetoplasmons in continuous 2DESs. Moreover, the interplay between a quantizing magnetic field and inter-disk Coulomb interactions is expected to yield novel magnetoplasmon band structures, potentially giving rise to topological edge states or magnetic-field-controlled slow-light effects~\cite{jin2019topological,jin2016topological}.

Gallium arsenide (GaAs) quantum wells provide an ideal platform for such studies due to their exceptionally high electron mobility, tunable density, and well-understood confinement potential. The ability to process high-quality 2DES disks via photolithography and etching allows precise control over both the disk radius and the lattice constant, making it possible to systematically vary the coupling strength. Despite this potential, a comprehensive experimental and theoretical investigation of how the magnetoplasmon dispersion evolves from isolated disks to a strongly coupled lattice remains lacking.

Key open questions include: How does the magnetic field dependence of the collective modes change as coupling increases? When does independent-disk approximation stop working?

In this paper, we address these questions by performing magneto-optical THz transmission measurements on a square lattices of GaAs 2DES disks with varying inter-disk distances. We map the magnetic-field-dependent resonance frequencies as a function of lattice constant and compare our experimental results with a non-interacting disks regime. These results advance the understanding of collective magnetoplasmonics in patterned semiconductor systems and open avenues for magnetic field-tunable plasmonic crystals and metasurfaces.

\section{Experimental technique}

The studies were carried out on a series of samples with a high-quality two-dimensional electron system based on an Al$_{0.3}$Ga$_{0.7}$As/GaAs/Al$_{0.3}$Ga$_{0.7}$As quantum well. Each structure consisted of a single quantum well $20$~nm wide, located at a depth of $200$~nm from the crystal surface. The two-dimensional electron concentration in the quantum well was $n_s = 1.1 \times 10^{12}$~cm$^{-2}$, with a mobility of $\mu = 10^5$~cm$^2$/Vs at a temperature $T=5$ K.  
On the surface of each sample, on the quantum well side, an array of disks in the form of a two-dimensional square lattice with period $a$ was fabricated lithographically. The disk radius was the same for all samples and equal to $R=50$ $\mu$m. The lattice period varied and was equal to $a=300$, $200$, and $110$~$\mu$m.  
The samples were square-shaped crystal plates with $1 \times 1$~cm$^{2}$ size and were mounted on a copper diaphragm with a diameter of $8$~mm at the center of a superconducting solenoid inside a cryostat with optical windows. The temperature of the sample was maintained at a constant level of $5$~K and magnetic field was swept up to 1 T. Magnetic field was oriented perpendicular to the sample. To cover the frequency range from 75 to 400 GHz we used a microwave generator with a set of frequency extenders. The transmitted radiation was collected by a pyroelectric detector coupled with a lock-in amplifier.

To determine the plasma resonance frequency in an array of disks made from a two-dimensional electron system, it is necessary to be able to suppress its conductivity. One recently proposed convenient method is to suppress the conductivity of the 2DES by means of an external magnetic field perpendicular to the sample surface. Indeed, the components of the conductivity tensor in the Drude model are written as:

\begin{equation}
\begin{split}
& \sigma_{xx} = \frac{n_s e^2 \tau}{m^{\ast}} \frac{1 + i \omega \tau}{(1 + i \omega \tau)^2 + \omega_c^2 \tau^2}, \\ 
&\sigma_{xy} = \frac{n_s e^2 \tau}{m^{\ast}} \frac{\omega_c \tau}{(1 + i \omega \tau)^2 + \omega_c^2 \tau^2},  \\
& t=\frac{1+ Z_0 \sigma_{xx}/2}{(1+ Z_0 \sigma_{xx}/2)^2 + (Z_0 \sigma_{xy}/2)^2}.
\end{split}
\end{equation}

When $\omega_c \tau =(eB/m^{\ast}) \tau \gg 1$ and $\omega_c \gg \omega$, the transmittance of the two-dimensional electron system becomes $t=1$, and the 2DES no longer affects the amplitude. To estimate a sufficient magnetic field one should note that cyclotron slope in GaAs is $\partial F/\partial B = e/2\pi m^* = 418$ GHz/T. Therefore, by dividing the transmission through the sample to a transmission in a field B $= 1$ T, we can construct a normalized transmission spectrum in which the plasma resonance becomes clearly visible.

\section{Experimental results}

\begin{center}
	\begin{figure}[h!]               
		\centering                  
		\includegraphics[width=0.7\textwidth]{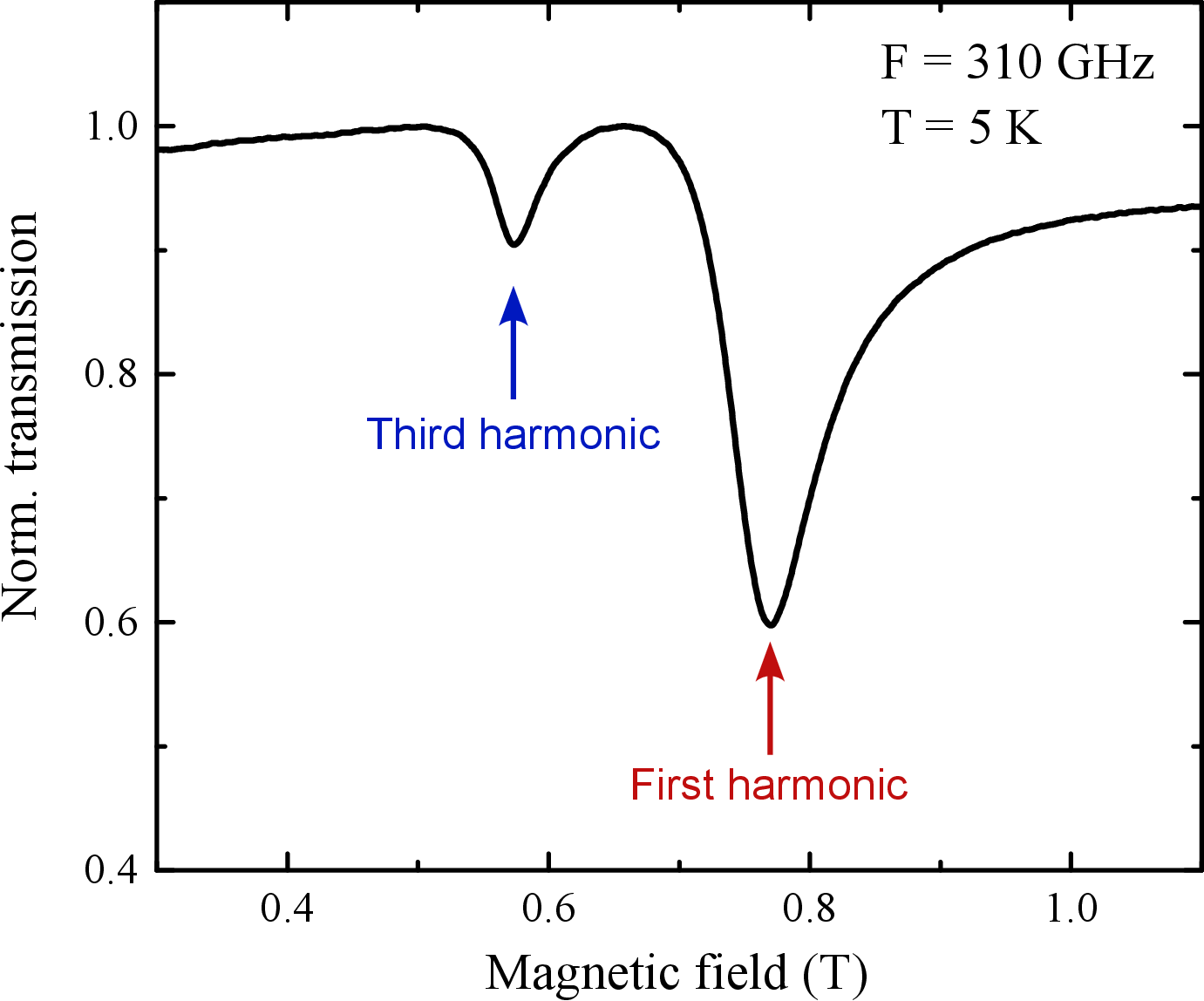} 
		\caption{Transmission dependence on magnetic field for a sample with $R$ = 50 and $a$ = 200 $\mu$m at a frequency F = 310 GHz and a temperature T = 5 K. The blue and red arrows correspond to a fundamental and a subsequent mode observed. Hereinafter red color illustrates first harmonic mode, while third harmonic is depicted in blue. Transmission is normalized to signal measured at B = 1 T for better feature visibility as described in the main text. }  
		\label{fig1}        
	\end{figure}
\end{center}

\begin{figure}[h!]               
	\centering                 
	\includegraphics[width=0.9\textwidth]{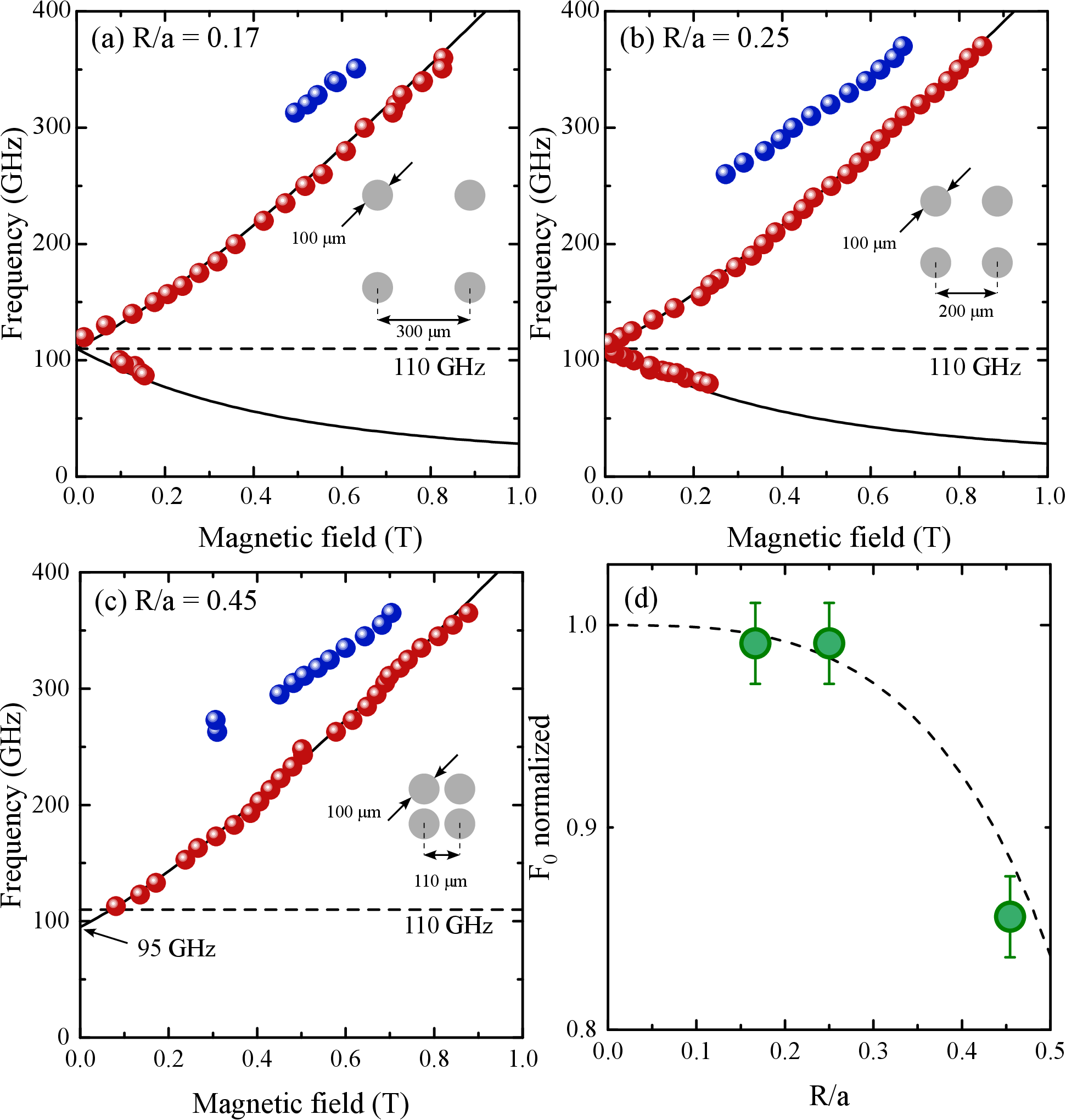} 
	\caption{Magnetodispersions for samples with $R$ = 50 $\mu$m and (a) $a$ = 300 $\mu$m, (b) 200 $\mu$m and (c) 110 $\mu$m measured at T = 5 K. Experimental data is illustrated with red and blue dots for first and third harmonic, respectively. Solid lines represent bulk and edge branches calculated according to Eq.~\ref{disp}. Horizontal dashed lines in panels (a)-(c) mark plasma frequency in zero magnetic field $F_0$ in the non-interacting disks regime. Insets demonstrate sample geometry for each case. Panel (c) has an additional arrow marking the $F_0$ being slightly decreased when disks are brought closer to one another; (d) Normalized plasma frequency dependence on $R/a$. The $F_0$ is normalized to its value in non-interacting disks case. Dashed line is a theoretical calculation according to~\cite{PhysRevB.54.10335}. } 
	\label{fig2}        
\end{figure}

We start by measuring a series of magnetic field sweeps while irradiating sample with a microwave radiation with fixed frequency. For a better feature visibility we plot transmission normalized to its value in magnetic field B $ = 1$ T. Typical transmission trace obtained in this case is shown in Fig.~\ref{fig1} for a sample  with $R$ = 50 and $a$ = 200 $\mu$m disks. There are two well-resolved resonance dips at approximately 0.57 and 0.78 T. To analyze these peaks, we calculate plasma frequency according to Eq.~\ref{2D}. For studied samples we substitute $m^*$ with $0.067  m_0$ for an effective mass in a GaAs, $\varepsilon_{GaAs} = 12.8$. The obtained plasma frequency value is 111 GHz. The magnetic field positions of resonant features in Fig.~\ref{fig1} allow us to identify these peaks as magnetoplasma harmonics and assign them numbers 1 and 3 according to Eq.~\ref{disp}, as their wave vector is equal to $q_{0}$ and $3q_{0}$, respectively. The more accurate expression for a wave vector could be obtained according to~\cite{PhysRevB.33.5221}. The vertical arrows in Fig.~\ref{fig1} indicate the resonances corresponding to the first and subsequent bulk magnetoplasma modes in a single disk. The absence of even-numbered plasma modes is not surprising, as it naturally reflects the distinct excitation conditions for modes of opposite parity under uniform electromagnetic radiation within the dipole approximation. Owing to symmetry, even-numbered modes possess zero dipole moment and thus remain optically inactive~\cite{PhysRevB.71.035320}, which is a typical case for plasmons with different parity~\cite{PhysRevB.73.113310}.

The main goal of this manuscript is devoted to uncover plasma frequency modification due to inter-disk distance. To do that, we repeat these magnetic field transmission measurements for all samples studied and do the same mode identification procedure. Therefore, we can plot these features on magnetodispersion and capture its behavior in dependence on distance between disks. Fig.~\ref{fig2} represent magnetodispersion for all samples in the study. Panels (a) to (c) correspond to disks with the same $R = 50$ $\mu $m, while disks lattice period varied and was equal to 300, 200, and 110 $\mu$m. Dispersion panels show visible first and third harmonics illustrated in red and blue colors, respectively. We also plot an edge mode measured for samples with $a$ = 300 and 200 $\mu$m. In case of the closest disks, edge and bulk modes in a low magnetic field are not well resolved due to lineshape widening possibly caused by disks interaction, hence in low frequency range those points are not plotted to the dispersion. 

The main feature in magnetoplasma dispersion graph is a degenerate point, where two branches coincide. We experimentally show that zero-field plasma frequency $F_0 = 110$ GHz for Fig.~\ref{fig2}(a) and (b) cases, which is in good agreement with a value calculated with Eq.~\ref{2D} being equal to 111 GHz. When disks are brought closer together, plasma frequency shift to the lower values and is equal to 95 GHz for a sample with $R/a = 0.45$.

To summarize our main result we plot a plasma frequency $F_0$ dependence on disk proximity factor $R/a$ in Fig.~\ref{fig2}(d). To verify our experimental data we also plot theoretical dependence according to the Eq.~23 from~\cite{PhysRevB.54.10335}:

\begin{equation}
    \dfrac{F_0^2}{F_0^2(R/a = 0)} = 1-\dfrac{2\eta(3/2)}{3\pi}\left(\dfrac{R}{a}\right)^3 - \dfrac{6\eta(5/2)}{5\pi}\left(\dfrac{R}{a}\right)^5+...,
\end{equation}

where $\eta(z) = \Sigma (k^2+l^2)^{-z}$ (the sum is taken over all $k$, $l$, excluding $k = l = 0$); $\eta(3/2) = 9.03$, $\eta(5/2) = 5.09$.

Although, a reduction in $F_0$ due to lateral disk screening was anticipated, both experiment and theory are in excellent agreement and demonstrate that this effect is influencing non-interacting $F_0$ value by approximately 15 percent. In conclusion, the system can be validly considered non-interacting as long as the disks are not brought into extremely close proximity; even then, inter-disk interactions introduce only a modest modification.

\section{Conclusion}
In this work, terahertz spectroscopy is utilized to experimentally investigate the plasmon resonance in an array of two-dimensional disks containing an electron system. From the measured transmission spectra, the frequencies of the fundamental mode are determined for two distinct regimes: $R/a \ll 0.5$ and $R/a \rightarrow 0.5$. It is established that, in the former case, the plasma frequency remains unaffected by lateral screening. The resonance frequency extracted from the magnetodispersion, being equal to $110$ GHz, is in excellent agreement with the non-retarding plasmon dispersion of an isolated disk. The plasma frequency and magnetodispersion are also measured in the regime $R/a \rightarrow 0.5$. As a result of inter-disk coupling, the plasma frequency decreases to $95$ GHz. Thus, even at small inter-disk separations, lateral screening in a disk array produces only mild effect. These findings may prove valuable for controlling collective effects in the fabrication of samples for terahertz spectroscopy.

\section{Acknowledgments}

The authors thank A.~A.~Zabolotnykh and D.~A.~Rodionov for fruitful discussions and encouraging feedback.

\bibliography{GaAs-bib}
\end{document}